\documentclass[prd,aps,floats,preprintnumbers,preprint]{revtex4}

\usepackage{graphicx, epsfig}
 
\newcommand{\be}{\begin{equation}}
\newcommand{\ee}{\end{equation}}
\newcommand{\bea}{\begin{eqnarray}}
\newcommand{\eea}{\end{eqnarray}}

\begin{document}


\setlength{\unitlength}{1mm}

\title{LTB  universes as alternatives to dark energy: does positive averaged acceleration imply positive cosmic acceleration? }

\author{Antonio Enea Romano}

\affiliation{
Department of Physics, University of Wisconsin, Madison, WI 53706, USA 
}

\begin{abstract}
We show that positive averaged acceleration $a_D$ obtained in LTB models through spatial averaging can require integration over a region beyond the event horizon of the central observer.
We provide an example of a LTB model with positive $a_D$ in which the luminosity distance $D_L(z)$ does not contain information about the entire spatially averaged region, making $a_D$ unobservable.
Since the cosmic acceleration $a^{FRW}$  is obtained from fitting the observed luminosity distance to a FRW model we conclude that in general a positive $a_D$ in LTB models does not imply a positive $a^{FRW}$. 
   
\end{abstract}

\maketitle
\section{Introduction}
High redshift luminosity distance measurements \cite{Perlmutter:1999np,Riess:1998cb,Tonry:2003zg,Knop:2003iy,Barris:2003dq,Riess:2004nr} and
WMAP measurement \cite{WMAP2003,Spergel:2006hy} of cosmic
microwave background (CMB) interpreted in the framework of standard FRW cosmological models have strongly disfavored a matter dominated universe, and strongly supported a dominant dark energy component, corresponding to a positive cosmological acceleration, which we will denote as $a^{FRW}$ (not to be confused with the scale factor $a$).
As an alternative to dark energy, it has been proposed \cite{Nambu:2005zn} that we may be at the center of a inhomogeneous isotropic universe described by a Lemaitre-Tolman-Bondi (LTB) solution of Einstein's fields equations, where spatial averaging over one expanding and one contracting region is producing a positive averaged acceleration $a_D$.
Another more general approach to map luminosity distance as a function of redshift $D_L(z)$ to LTB models has been recently proposed  \cite{Chung:2006xh}, showing that an inversion method can be applied successfully to reproduce the observed $D_L(z)$.   

The main point is that the luminosity distance is in general sensitive to the geometry of the space through which photons are propagating along light geodesics, and therefore arranging appropriately the geometry of a given cosmological model it is possible to reproduce a given $D_L(z)$.
For FRW models this correspond to set constraints on    $\Omega_{\Lambda}$ and $\Omega_m$ and for LTB models it allows to determine the functions $E(r),M(r),t_b(r)$.

The averaged acceleration $a_D$ on the other side is not directly related to $a^{FRW}$, since this is obtained integrating the position dependent cosmological redshift along the light geodesics, while $a_D$ is the result of spatial averaging, and has no relation to the causal structure of the underlying space.
This crucial difference can make $a_D$ unobservable to a central observer $O_c$ when the scale of the spatial averaging is greater than its event horizon.
  
\section{Lemaitre-Tolman-Bondi (LTB) Solution\label{ltb}}
Lemaitre-Tolman-Bondi  solution can be written as \cite{Lemaitre:1933qe,Tolman:1934za,Bondi:1947av} as:
\begin{eqnarray}
\label{eq1} %
ds^2 = -dt^2  + \frac{\left(R,_{r}\right)^2 dr^2}{1 + 2\,E(r)}+R^2
d\Omega^2 \, ,
\end{eqnarray}
where $R$ is a function of the time coordinate $t$ and the radial
coordinate $r$, $E(r)$ is an arbitrary function of $r$, and
$R,_{r}$ denotes the partial derivative of $R$ with respect to
$r$.

Einstein's equations give:
\begin{eqnarray}
\label{eq2} \left({\frac{\dot{R}}{R}}\right)^2&=&\frac{2
E(r)}{R^2}+\frac{2M(r)}{R^3} \, , \\
\label{eq3} \rho(t,r)&=&\frac{M,_{r}}{R^2 R,_{r}} \, ,
\end{eqnarray}
with $M(r)$ being an arbitrary function of $r$ and the dot
denoting the partial derivative with respect to $t$. The solution
of Eq.\ (\ref{eq2}) can be written parametrically by using a
variable $\eta=\int dt/R \,$, as follows
\begin{eqnarray}
\label{eq4} \tilde{R}(\eta ,r) &=& \frac{M(r)}{- 2 E(r)}
     \left[ 1 - \cos \left(\sqrt{-2 E(r)} \eta \right) \right] \, ,\\
\label{eq5} t(\eta ,r) &=& \frac{M(r)}{- 2 E(r)}
     \left[ \eta -\frac{1}{\sqrt{-2 E(r)} } \sin \left(\sqrt{-2 E(r)}
     \eta \right) \right] + t_{b}(r) \, ,
\end{eqnarray}
where  $\tilde{R}$ has been introduced to make clear the distinction between the two functions $R(t,r)$ and $\tilde{R}(\eta,r)$ which are trivially related by 

\begin{equation}
R(t,r)=\tilde{R}(\eta(t,r),r)
\label{Rtilde}
\end{equation}

and $t_{b}(r)$ is another arbitrary function of $r$, called bang function, which corresponds to the fact that big-bang/crunches happen at different times in this space. This inhomogeneity of the location of the singularities is the origin of the possible causal separation between the central observer and the spatially averaged region for models with positive $a_V$.

We can introduce the following variables
\begin{equation}
 a(t,r)=\frac{R(t,r)}{r},\quad k(r)=-\frac{2E(r)}{r^2},\quad
  \rho_0(r)=\frac{6M(r)}{r^3} \, ,
\end{equation}
so that  Eq.\ (\ref{eq1}) and the Einstein equations
(\ref{eq2}) and (\ref{eq3}) can be written in a form which is more similar to FRW models

\begin{equation}
\label{eq6} ds^2 =
-dt^2+a^2\left[\left(1+\frac{a,_{r}r}{a}\right)^2
    \frac{dr^2}{1-k(r)r^2}+r^2d\Omega_2^2\right] \, ,
\end{equation}
\begin{eqnarray}
\label{eq7} %
\left(\frac{\dot{a}}{a}\right)^2 &=&
-\frac{k(r)}{a^2}+\frac{\rho_0(r)}{3a^3} \, ,\\
\label{eq:LTB rho 2} %
\rho(t,r) &=& \frac{(\rho_0 r^3)_{, r}}{6 a^2 r^2 (ar)_{, r}} \, .
\end{eqnarray}
The solution in Eqs.\ (\ref{eq4}) and (\ref{eq5}) can now be written as
\begin{eqnarray}
\label{LTB soln2 R} \tilde{a}(\tilde{\eta},r) &=& \frac{\rho_0(r)}{6k(r)}
     \left[ 1 - \cos \left( \sqrt{k(r)} \, \tilde{\eta} \right) \right] \, ,\\
\label{LTB soln2 t} t(\tilde{\eta},r) &=& \frac{\rho_0(r)}{6k(r)}
     \left[ \tilde{\eta} -\frac{1}{\sqrt{k(r)}} \sin
     \left(\sqrt{k(r)} \, \tilde{\eta} \right) \right] + t_{b}(r) \, ,
\end{eqnarray}
where $\tilde{\eta} \equiv \eta r = \int dt/a \,$.

\section{Averaged acceleration }


Following the standard averaging procedure \cite{Buchert:1999mc,Palle:2002zf,Kolb:2005da,Nambu:2005zn} we define the volume for a spherical domain, $0<r<r_D$, as 
\begin{equation} \label{eq:ltb volume}
V_D %
= 4 \pi \int_0^{r_D} \frac{R^2 R,_{r}}{\sqrt{1+2E(r)}} \, dr 
\end{equation}
and the length associated to the domain as
\
\begin{equation}
\label{LD}
L_D = V_D^{1/3},	
\end{equation}

via which the deceleration
parameter $q_D$ and the average acceleration $a_D$ (not to be confused with the scale factor $a$) are defined as:

\begin{eqnarray}
\label{qD}
q_D & = & -\ddot{L}_D L_D/\dot{L}_D^2 \\
\label{aD}
a_D & = & \dot{L}/L	
\end{eqnarray}

The models studied in \cite{Chuang:2005yi} correspond to the following functions $k(r),\rho_0(r),t_b(r)$:

\begin{eqnarray}
t_b(r) &=& - \frac{h_{tb}(r/r_t)^{n_t}}{1+(r/r_t)^{n_t}}    \, ,  \\
k(r) &=& -\frac{(h_k+1) (r/r_k)^{n_k}}{1+(r/r_k)^{n_k}}+1 \, ,  \\
\rho_0 (r) &=& \textrm{constant}  \, .
\end{eqnarray}
After exploring the 9 parameters space they give three examples of LTB solutions with positive $q_D$ as shown in Table I.

\tabcolsep=8pt
\begin{table}[ht!]
\caption{Three examples of the domain acceleration.} \label{table:domain eg} %
\center
\begin{tabular}{|c|c|c|c|c|c|c|c|c|c||c|}\hline
&$t$ & $r_D$ & $\rho_0$ & $r_k$ & $n_k$ & $h_k$ & $r_t$ & $n_t$ & $h_{tb}$ & $q_D$ \\
\hline
1&$0.1$ & $1$ & $1$ & $0.6$ & $20$ & $10$ & $0.6$ & $20$ & $10$ & $-0.0108$ \\
\hline
2&$0.1$ & $1.1$ & $10^5$ & $0.9$ & $40$ & $40$ & $0.9$ & $40$ & $10$ & $-1.08$ \\
\hline
3&$10^{-8}$ & $1$ & $10^{10}$ & $0.77$ & $100$ & $100$ & $0.92$ & $100$ & $50$ & $-6.35$ \\
\hline
\end{tabular}

\begin{tabular}{|c|c|c|c|c|}
\hline
& $L_D$ & $\dot{L}_D$ & $\ddot{L}_D$ & $q_D$\\
\hline
1&$16.2$&$1.62$&$0.00174$& $-0.0108$\\
\hline
2& $94.0$ & $7.63$ & $0.694$ & $-1.08$ \\
\hline
3&$8720$&$117$&$10.0$& $-6.35$\\
\hline
\end{tabular}
\end{table}

\begin{figure}[h]
	\begin{center}
		\includegraphics{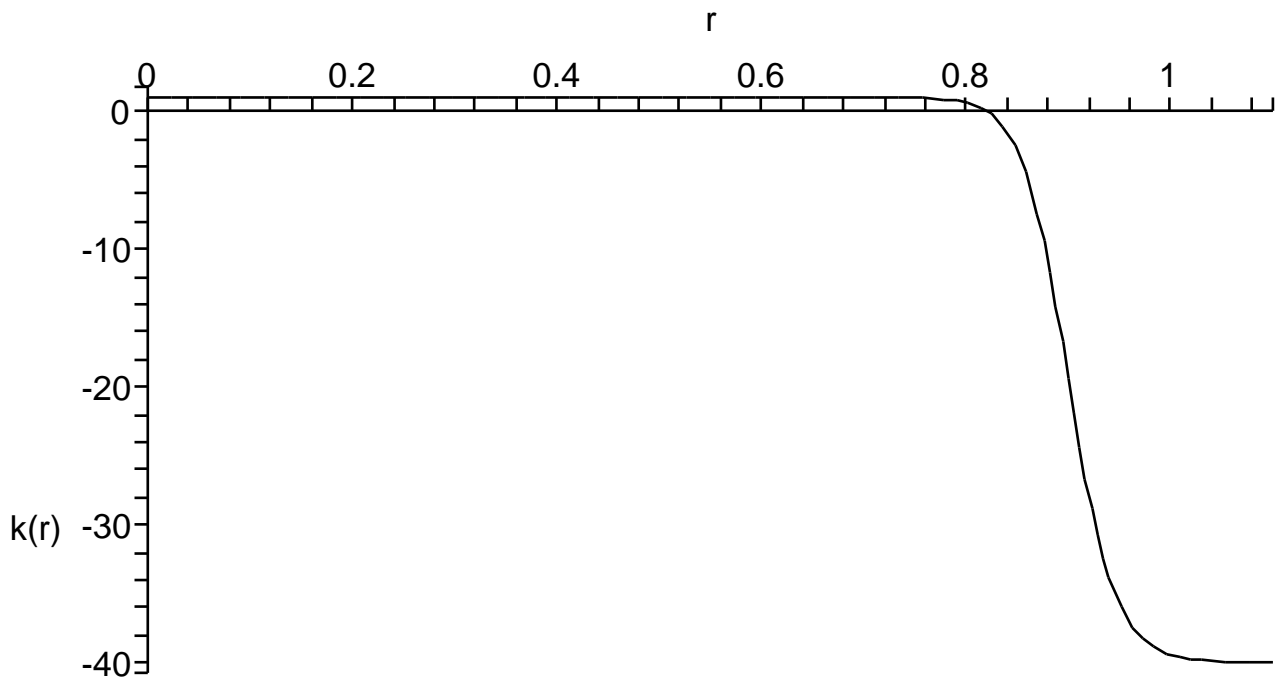}
		\end{center}	
		\caption{$k(r)$ is plotted for the model corresponding to row 2 of Table I from \cite{Chuang:2005yi}. }
	\label{fig:k(r)}
\end{figure}

Since $t(0,r)=t_b(r)$ and $\tilde{R}(0,r)=0$
we can fix the following initial condition for $R(t,r)$:
\begin{equation}
\label{INI0}
\tilde{R}(\eta=0,r)=0=R(t(\eta=0,r),r)=R(t_b(r),r) 
\end{equation}
which clearly shows why $t_b(r)$ is called bang function.

Defining $t_q$ as the time in the first column of Table I, i.e. the time at which $q(t_q)=q_D$, with $q_D$ being the value in the last column of the same table, we solved the light geodesics equation 

\begin{eqnarray}
\frac{dT(r)}{dr}=-\,{R'(r,T(r))\over {\sqrt{1+2E(r)}}}. \label{eq:27}
\end{eqnarray}

imposing the following initial conditions 
\begin{eqnarray}
\label{INI1}	T(r=0) & = &t_q  \\
\label{INI2}  R(t_b(r),r)& = & 0 
\end{eqnarray}

Eq.({\ref{INI1}}) is  the natural way to map these models into luminosity distance observations for a central observer which should receive the light rays at the time $t_q$ at which the averaged acceleration is positive. 
In other words the time dependence of $a_D(t)$ corresponds to the choice of initial conditions for the light geodesics used to compute the luminosity distance.    

Solving the equation

\begin{equation}
\label{Hor} 
T(r_{Hor})=t_b(r_{Hor})
\end{equation}
for the model corresponding to the second row of Table I we obtain $r_{Hor}<r_D$, where $r_D$ is the upper limit of the volume averaging integral. In this context $r_{Hor}$ can be thought of as the comoving horizon, the maximum radial coordinate from which photons can reach the central observer $O_c$ at time $t_q$.

\begin{figure}[h]
	\begin{center}
		\includegraphics{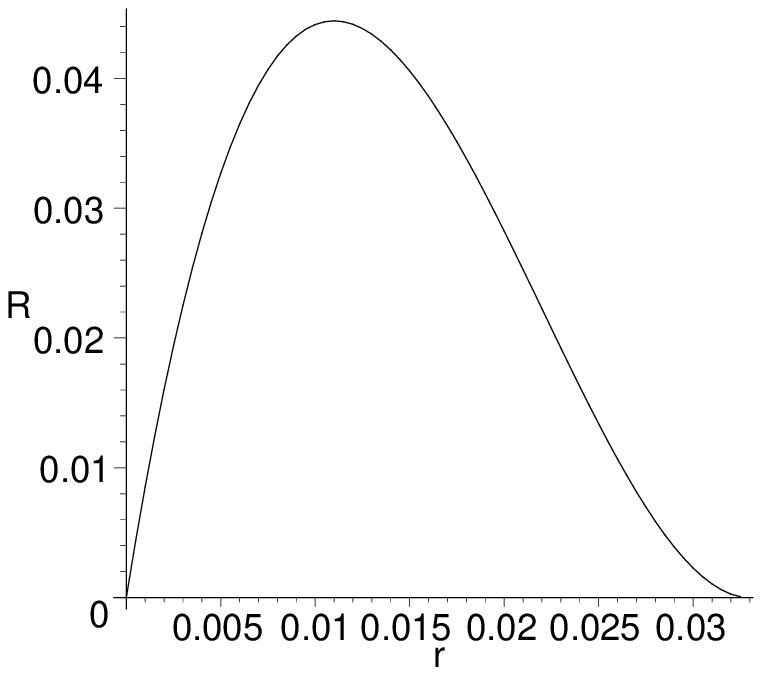}
		\includegraphics{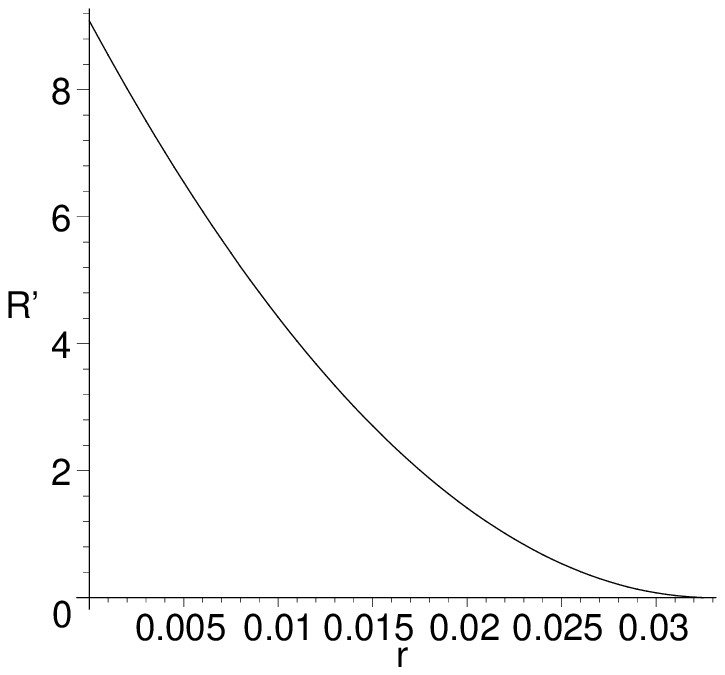}
	\end{center}	
		\caption{$R(T(r),r)$ and $R'(T(r),r)$ are plotted for model 2 of \cite{Chuang:2005yi}.At about $r_{Hor}$=0.03298 there is a singularity. }
	\label{fig:m2RRr}
\end{figure}

As it can be seen in Fig. \ref{fig:m2RRr} at  $r=r_{Hor}$ there is a singularity along the geodesic, corresponding to the fact that light cannot reach the central observer at time $t_q$ from points at radial coordinate $r>r_{Hor}$.
In fact, since $t_b(r)$ is a decreasing function of r, photons emitted at $r>r_{Hor}$ will arrive at $r=r_{Hor}$ at time $T(r_{Hor})$ when a local singularity develops because
\begin{equation}
\label{sing}
 R(T(r_{Hor}),r_{Hor})=R(t_b(r_{Hor}),r_{Hor})=0.  
\end{equation}
Since regions of the universe at radial coordinates greater than $r_{Hor}$ have never been in causal contact with $O_c$ at time $t_q$, the scale at which $q_D$ is defined is beyond the region causally connected to $O_c$. Therefore $q_D$ cannot be detected from local observation of the luminosity distance which is used to define $a^{FRW}$.
This result is not surprising since the spatial averaging procedure is insensitive to the causal structure of the underlying space.
The other models studied in \cite{Chuang:2005yi} do not show this causal behavior, but solving the geodesic equations and computing the luminosity distance using the same initial conditions adopted above it can be shown that they do not reproduce correctly the observed luminosity distance, but this goes beyond the scope of the present article, which is the study of the local observability of the averaged acceleration.
We will study the luminosity distance of these and other LTB models with positive $a_D$, and the relation between $q_D$ and $q^{FRW}$ in another work.

\begin{figure}[t]
	\begin{center}
	\includegraphics{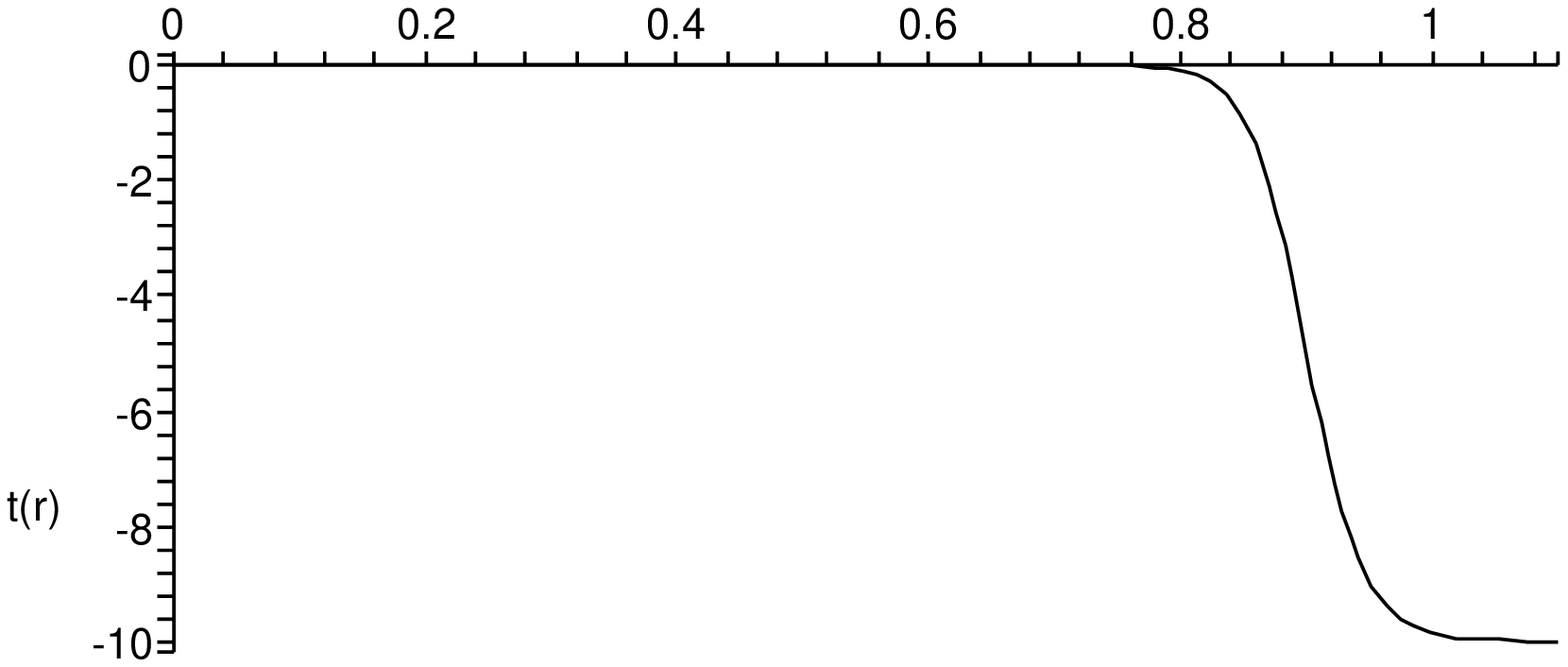}
	\end{center}	
		\caption{$t_b(r)$ is plotted for the model corresponding to row 2 of Table I from \cite{Chuang:2005yi}. }
	\label{fig:tb(r)}
\end{figure}

\begin{figure}[t]
	\begin{center}
		\includegraphics{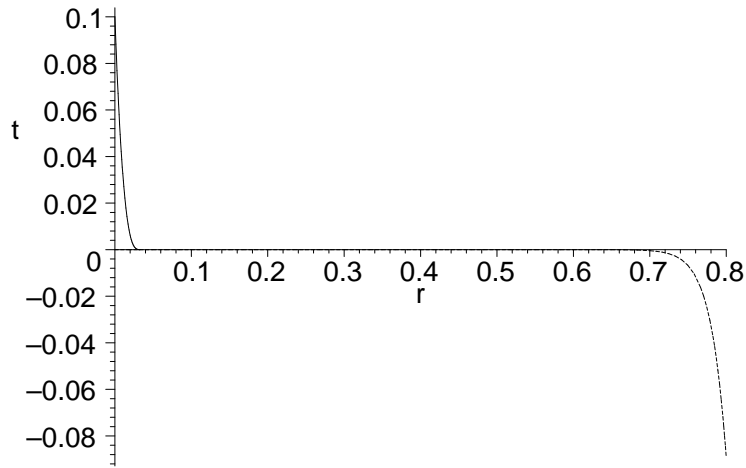}
	\end{center}
		\caption{The solid curve shows the light geodesic $T(r)$ and the dashed curve shows the bang function $t_b(r)$. For $r<0.6$  $t_b(r)$ is so small that it cannot be distinguished from the horizontal zero axis.}
	\label{fig:m2tg-tb}
\end{figure}

This example shows how $a_D$ may not even be causally related to the local observation of $D_L(z)$, and gives a reverse example of the results obtained in \cite{Enqvist:2006cg}, where it was studied a LTB model fitting the observed luminosity distance, consistent with a positive $a^{FRW}$, but without positive averaged acceleration $a_D$.
Our results do not rule out LTB models as alternatives to dark energy since, since  the inversion method \cite{Chung:2006xh} allows to obtain the observed luminosity distance without any averaging, and some concrete examples derived independently have already been proposed \cite{Alnes:2006uk,Alnes:2005rw,Enqvist:2006cg}.

We can conclude that the luminosity distance $D_L(z)$ contains more information than the spatially averaged acceleration $a_D$ because the first is sensitive to the causal structure of the entire space-time while the second is the result of averaging only the spatial part of the geometry, making the relation between them in general not one-to-one. 
Our results confirm the general argument \cite{Paranjape:2006ww} that the spatial averaging must be performed on comoving domains whose size coincides with the length scale at which homogeneity sets in, and not on domains which may be as large as or larger than the observable universe. 
It is possible that writing the geodesics equation in a form consistent with the Buchert averaged formalism could overcome the problem, but this would be just a formal solution, since light is propagating along the geodesics of the inhomogeneous space, not of the averaged one.
\section{Discussion}

We showed that LTB models with positive averaged acceleration can require averaging on scales beyond the event horizon of the central observer. In these cases $a_D$ is causally unrelated to $a^{FRW}$ which is obtained from the observed luminosity distance, showing that a positive averaged acceleration $a_D$ in LTB models is in general not equivalent to a positive $a^{FRW}$.
This example shows how the averaging Buchert \cite{Buchert:1999mc} formalism is not preserving the causal structure of space-time and can lead to the definition of locally unobservable averaged quantities.
The study of the constraints on the local observability of averaged quantities will be addressed in a more general way, not only in the context of LTB models, in a future work.


\begin{acknowledgments}
I thank E. Kolb, D. Chung and  A. Paranjape for useful comments and discussions, and IAC for the hospitality.
I also thank Desiree for her moral support, and Mariuccia for her kind hospitality.

\end{acknowledgments}

\end{document}